# Explaining temporal trends in annualized relapse rates in placebo groups of randomized controlled trials in relapsing multiple sclerosis: systematic review and meta-regression

**Simon M Steinvorth [i], Christian Röver [i], Simon Schneider [i], Richard Nicholas [ii], Sebastian Straube [iii], Tim Friede [i]**

[i] Department of Medical Statistics, University Medical Center Göttingen, Göttingen, Germany

[ii] Imperial College Healthcare NHS Trust, London, UK

[iii] Department of Occupational, Social and Environmental Medicine, University Medical Center Göttingen, Göttingen, Germany

*Corresponding author:*

Prof Tim Friede, University Medical Center Göttingen, Department of Medical Statistics, Humboldtallee 32, D-37073 Göttingen; Tel +49-551-39 4991; Fax +49-551-39 4995; tim.friede@med.uni-goettingen.de

**Abstract**

**Background:** Recent studies have shown a decrease in annualized relapse rates (ARRs) in placebo groups of randomized controlled trials (RCTs) in relapsing multiple sclerosis (RMS).

**Methods:** We conducted a systematic literature search of RCTs in RMS. Data on eligibility criteria and baseline characteristics were extracted and tested for significant trends over time. A meta-regression was conducted to estimate their contribution to the decrease of trial ARRs over time.

**Results:** 56 studies were identified. Patient age at baseline ($p<0.001$), mean duration of MS at baseline ($p=0.048$), size of treatment groups ($p=0.003$), Oxford Quality Scale scores ($p=0.021$), and the number of eligibility criteria increased significantly ($p<0.001$), whereas pre-trial ARR ($p=0.001$), the time span over which pre-trial ARR was calculated ($p<0.001$), and the duration of placebo-controlled follow-up ($p=0.006$) decreased significantly over time. In meta-regression of trial placebo ARR the temporal trend was found to be insignificant, with major factors explaining the variation being pre-trial ARR, the number of years used to calculate pre-trial ARR, and study duration.

**Conclusion:** The observed decline in trial ARRs may result from decreasing pre-trial ARRs and a shorter time period over which pre-trial ARRs were calculated. Increasing patient age and duration of illness may also contribute.

**Introduction**

Relapsing MS (RMS) is defined by the presence of relapses, a neurological deterioration lasting greater than 24 hours with stabilization or recovery [1]. Relapses occur in relapsing-remitting MS (RRMS), secondary progressive MS (SPMS) and progressive-relapsing MS (PRMS) [2]. The development of therapies that have successfully targeted relapses has meant that relapse focused outcome measures are common in RMS trials – of particular importance is the annualized relapse rate (ARR) [3, 4]. Recently, a downward trend in trial ARRs of placebo patients has been identified in randomized controlled trials (RCTs) [5, 6]. Although placebo patients receive no active agent, their ARRs improve significantly compared to baseline and increasingly so [7]. This trend is unexplained but has implications for inter-trial comparability of ARRs [6]. This is particularly important as more therapies are licensed in the absence of head-to-head data. We investigated possible reasons for this downward trend in trial placebo ARR over time by analyzing patient baseline characteristics, eligibility criteria and study design features in RCTs.

**Methods**

*Systematic literature search*

A systematic literature search was conducted in PubMed, the Cochrane Library, and Web of Science aiming to identify placebo-controlled, double-blind RCTs in MS with data on trial ARR or, alternatively, data that allowed trial ARR to be calculated. We searched [July 13, 2012] using combinations of the search terms „multiple sclerosis", „ms", „placebo", „controlled", „control", „relapsing", „remitting", „relapse", „exacerbation", „lesion", „reduction", „rate", „patients", „double-blind", and „clinical". (see supplementary material). We excluded cross-over trials and studies where patients in the control group received add-on therapies.

*Data extraction*

The following data were extracted by one reviewer and verified by another:

- the trial ARR of the placebo group along with the associated number of placebo patients and the time period over which the ARR was calculated;
- the number of eligibility criteria, words and characters used to describe the criteria;
- trial inclusion criteria: age, pre-trial ARR and Expanded Disability Status Scale (EDSS) score at baseline;
- trial inclusion criteria describing the minimum number of days since the last relapse or use of high-dose steroids;
- patient baseline characteristics: age, pre-trial ARR, EDSS, and duration of MS;
- the number of patients and the number of female patients;
- the Oxford Quality Scale (OQS) score was calculated [8].

If possible, mean values were extracted along with standard deviations, or median values and interquartile ranges. Where standard deviations were not available, they were calculated from p-values, standard errors, confidence intervals or t-statistics [9]. If an ARR was not stated, it was calculated by dividing the total number of relapses by the number of patients in the placebo group, giving a mean relapse rate, and then extrapolating to an annualized relapse rate by correcting for the time over which relapses were observed. When an adjusted rate was given, i.e. adjusted to age, sex or other parameters, as well as an unadjusted rate, the latter was preferred. Whenever trials distinguished between different intensities of relapses, the total sum of relapses was counted irrespective of severity. For ARRs without a quoted standard error we derived errors based on a Poisson approximation.

For our count of eligibility criteria, all inclusion criteria counted, unless they were mutually exclusive, and all exclusion criteria were counted, unless they matched any inclusion criteria already counted, similar to the approach established by Clisant et al. [10]. Only criteria applying to patients with RMS counted. Having the correct diagnosis to be included to a study counted as one criterion. To determine the number of words or characters used to describe the eligibility criteria, all sentences or tables containing these criteria were copied into a text processing program (LibreOffice 3.5.2.2) and counted automatically. Features due to editing such as spaces and bullets were not included in the count. Captions of inclusion or exclusion criteria did count. When supplementary material offered more detailed information on eligibility criteria than the main publication, it was used instead of the latter.

Baseline characteristics of patients with RMS for the placebo group and all patients were retrieved. Where possible, data on RRMS patients were preferred over data on patients with other forms of MS. When the mean age at baseline was not given, it was calculated by adding the mean MS duration to the mean age at the onset of the disease, if provided. When pre-trial ARR was not specified, but the number of relapses in a certain time period or a non-annualized relapse rate, it was calculated as above. Whenever a study presented multiple pre-trial ARRs, calculated over different time periods, all were extracted. When baseline characteristics for the total patient group were not available, they were calculated by combining data provided for the individual treatment arms. Baseline characteristics of patients randomized to treatment arms were preferred over characteristics describing only patients that actually received treatment.

When multiple pre-trial ARRs could be collected, those accounting for the longest period were used. In these cases, pre-trial ARRs of the same groups calculated over different time periods were compared. When inclusion criteria appeared in a complex context, allowing alternative options to qualify in one measure or mutually exclusive options of different measures, items were omitted.

*Data analysis*

For the purpose of all analyses of temporal trends, the year and month of publication was used. When mean values were not given, available median values were used as a direct estimate instead, if they did not require further mathematical handling. Corresponding interquartile ranges, if not equal to zero, were used to estimate standard deviations, assuming normal distributions. Values obtained in such a manner were indicated in the figures. Mean values with standard deviations had top priority, mean values with standard deviations estimated from interquartile ranges second, median values with standard deviations estimated from interquartile ranges third priority, followed by solitary mean values and lastly solitary median values.

The natural logarithms of the trial and pre-trial ARR of placebo groups were modeled by Gaussian linear regression weighted by the inverse standard error squared. For the predicted means 95% confidence intervals were calculated. For baseline characteristics (age, disease duration, EDSS score, gender distribution) we calculated linear regressions over time, weighted by the inverse standard error squared, taking all values with standard errors into account. We calculated (unweighted) linear regressions for the number of eligibility criteria, the number of words or characters used to describe these criteria, the minimum pre-trial ARR for inclusion, the number of years over which the pre-trial ARR was calculated, the minimum number of days before baseline without the use of high dose steroids, the minimum number of days without relapse, the duration of placebo-controlled follow-up in days, the number of treatment arms, the average number of patients in each treatment arm, the score on the Oxford Quality Scale, and the patient years considered in its calculation.

As supportive analyses, we compared the mean values of four deliberate partitions of the 56 trials included in this study, testing for possible trends that might have been concealed in the analysis of all studies over time. We decided to subdivide the trials at recognizable points in trial history: The first cluster of trials comprised all trials up to the end of 1994, the second cluster all trials from the beginning of 1995 to the end of 2000, the third cluster all trials from the beginning of 2001 to the end of 2009, the fourth cluster all trials from the beginning of 2010 to today. Below we refer to these analyses as epoch analyses.

The logarithmic ratios of ARRs comparing the different pre-trial time periods were investigated via a random-effects meta-analysis with inverse variance weighting. The combined estimates are reported along with 95% confidence intervals (CIs) and p-values testing the null hypothesis of no difference between the time intervals. Heterogeneity between studies is estimated and reported in terms of the heterogeneity measure $I^2$, which is the ratio of the between-trial variance and the total variance, alongside the p-values of the chi-square test of heterogeneity. Forest plots illustrating the ratios of the individual studies and the combined effect allow for visual comparison of the heterogeneity and provide an overview of the results. The meta-analysis was conducted using the RevMan 5.1 software (http://ims.cochrane.org/revman).

Finally, all statistically significant temporal trends were investigated in a meta-regression calculating to what extent they contributed to the temporal trend in trial ARRs. The final combination of included variables was chosen in respect to the Bayesian information criterion (BIC), allowing a model as simple as possible with as little as four variables, yet explaining the bulk of the downward trend in trial ARR. In all instances, the level of statistical significance was set at 5%.

**Results**

A total of 56 randomized, placebo-controlled trials (references given in the web appendix) were identified including 14,792 patients of which 5,380 had been randomized to placebo. The study duration was on average 12 months (range 3 to 60 months). The total follow-up was 23,157 patient-years including 8,696 patient-years on placebo. Table 1 gives an overview of the patient populations included in the 56 trials and further information is included in the web appendix (supplementary material). Trial ARRs in placebo patients decreased (Figure 1) by 4.5% per year (95% CI (3.2; 5.7); p<0.001). This confirms previous findings by Inusah et al. [5] and Nicholas et al. [6] who conducted their reviews on different although overlapping sets of studies.

*Temporal trends in pre-trial ARR and associated factors*

Pre-trial ARR decreased over the last 30 years by 2.0% per year (95% CI (1.3; 2.6); p<0.001, Figure 1). From 1982 to 2012, the number of years over which pre-trial ARR was calculated, decreased significantly by approximately 1.5 years (p<0.001). The minimum pre-trial ARR for inclusion did not change significantly over time. The required days prior to baseline without relapse or high-dose steroid use likewise did not change. Seven studies provided data on pre-trial ARRs calculated over one and two years (Figure 2). The ARRs in the second year before inclusion (months -13 to months -24) were reduced by 49% (95% CI (44%; 54%), p<0.001) compared to the year prior to the study (months -1 to -12).

*Changes in other baseline characteristics over time*

We detected statistically significant trends over time in the following baseline characteristics. Baseline age increased by approximately one additional year for every five years (p<0.001), as did baseline MS duration, which increased by one additional year of MS duration every eight years (p=0.048, Figure 3). Baseline EDSS scores and minimum steroid-free time before inclusion in the trial were not found to be significant in the trend analyses (p=0.289 and p=0.059, respectively), but changes over time were found in the epoch analyses (p=0.003 and p=0.02, respectively). The mean minimum number of steroid-free days is 68 days for the first cluster, 35 days for the second, 38 days for the third, and 32 days for the fourth cluster of trials. The mean baseline scores on the EDSS are 3.67 for the first cluster of trials, 2.47 for the second, 2.35 for the third, and 2.72 for the fourth. The percentages of female patients in trials did not change over time.

*Changing eligibility criteria and study design characteristics over time*

The number of eligibility criteria increased significantly by one criterion every 16 months (p<0.001). The number of words used in the description of these criteria increased by about 7 words per year (p<0.001), as did the number of characters (about 40 characters per year, p<0.001). The duration of the placebo-controlled follow-up changed significantly with a shortening by 16 days of follow-up per year (p=0.006) and the number of patients in the placebo groups increased by 6.7 patients per year (p=0.004) on average. The number of patient–years (sample size × follow-up time) used to calculate the trial ARR increased, albeit non-significantly, over time with 8.5 additional patient years per annum (p=0.051). The number of treatment arms per trial increased significantly by nearly 1.5 over the last three decades (p<0.001), as did the average number of patients in each treatment arm (by 7.0 patients per year, p=0.003). Scores on the OQS slightly increased over time with, on average, studies scoring an additional point on the scale, which is scored out of five, every 36 years (p=0.021).

*Explaining temporal trends in trial ARR*

The temporal trend line for the trial ARR shown in Figure 1 explains about 46% of the variation observed in trial ARR over the years (Figure 4, left column). To gain insights into the drivers of this trend we utilized meta-regression incorporating changes in patient populations and trial characteristics. After taking all possible combinations of variables into consideration, we included pre-trial ARR, the number of years used to calculate pre-trial ARR, study duration and mean baseline MS duration. For comparison with the simple model including the time trend only we also added the year of publication. In the resulting model explaining about 69% of the variation in trial ARR the temporal trend is insignificant with major contributors being pre-trial ARR, the number of years used to calculate pre-trial ARR, study duration, and MS duration (Figure 4, right column).

**Discussion**

The aim of this study was to explain the decrease in trial ARRs in placebo patients that has been demonstrated previously [5, 6]. We found it to decrease by 4.5% each year in this study. One possible cause is the observed decrease in pre-trial ARR by nearly 0.8 relapses per year over the last three decades. This, in turn, may be related to the increasing age and duration of MS of patients in the trials. Very recently pre-trial ARR and mean baseline age were independently identified as predictors for on-trial ARR in a smaller number (n=13) of phase III trials with at least 18 months follow-up by Stellmann et al. [11]. We noted an increase in patients' average age over time; in our systematic review, patients were at baseline on average approximately six years older at the end of our period of observation compared to the beginning of our observation period. Tremlett et al. found a reduction in ARRs by 17% for every 5 years of MS duration [12]. The increase in MS duration by 3.6 years, as observed in our studies, could constitute a 13% decrease in ARRs. Considering this, in an older trial population with longer disease duration a decrease in pre-trial ARR with an associated decreasing trial ARR is to be expected. With EDSS scores remaining relatively stable after an early drop, especially in trials since 1995, this means that patients in newer studies tend to be later in their disease courses with less disability compared to patients in older studies. Thus they are likely to have less severe disease courses. However, one might speculate that longer disease durations might also reflect earlier diagnoses. The likely driver of these changes in study populations is the widening availability of increasingly effective treatment modifying those deemed suitable for trials by clinicians [5]. While the cause of temporal trends is most likely related to the patient recruitment period, this is not always available, and given the long-term nature of trends the publication dates should constitute a reasonable proxy. Besides changes in patient recruitment mechanisms, another potential factor might be the geographical region, but as studies are typically multi-center and often multi-national, this is among the questions that are beyond what can be addressed on the basis of the aggregate data.

Another factor contributing to the decrease in trial ARRs is the reduced time period over which pre-trial ARRs had been calculated; this decreased on average by 1.5 years over the past three decades. Using shorter periods of time over which pre-trial ARRs are calculated might thereby allow trials to include patients who, if pre-trial ARR was assessed over a longer time span, might not have been eligible for trial inclusion. We suspect the shortening of the time period considered for the estimation of the pre-trial ARR to be a principal factor driving the *regression to the mean* effect that was previously described by Martínez-Yélamos et al. [13] and Nicholas et al. [7], and that also appears to be evident in the findings of Kappos et al. [14]. The notion that patients are recruited into a study shortly after a flare up is supported by the finding that the ARR in the second year prior to recruitment is just half of the ARR in the year prior to the study. An increase in the number of phase III studies might also be a cause for the decrease in trial ARR – as phase III studies usually last longer than other studies, the longer duration may again contribute to a larger *regression to the mean* effect, and hence a larger discrepancy between trial and pre-trial ARR [7].

While growing numbers of eligibility criteria reflect the increasing understanding and complexity of possible influences on outcome variables such as the trial ARR, early trials with fewer eligibility criteria might have been more susceptible to such influences than modern ones. Similarly, changing definitions of MS and relapses, as well as varying forms of report, confirmation and treatment in case of relapses undoubtedly play a role, as has been suggested by Inusah et al. [7]. The Oxford Quality Scale scores of the trials in RMS generally increased over time – reflecting higher trial quality or better reporting as described [15].

Since the relative incidence of MS in women compared to men has risen from 2:1 to 4:1 over recent years [16], we were surprised to find stability in the proportion of patients who were women. Held et al. showed a correlation between on-study relapse rates and female sex [17] and therefore changes in the gender ratio over time could potentially explain part of the temporal trends in trial ARR.

According to our meta-regression the temporal trend observed by Inusah et al. [5] and Nicholas et al. [6] becomes relatively insignificant and other changes explain about 60% of the variation in the trial ARR. Similar analyses, aiming at modelling the pre-trial ARR, or excluding the pre-trial ARR from the set of predictors, yield comparable R² values using different sets of explanatory variables with age, number of eligibility criteria and number of patients being among the influential variables in addition. While the use of many partly correlated variables, the range of which is partly determined by what may be elicited from the investigated studies, and any model selection approach employed may have their issues, it is obvious that a range of possibly relevant factors are related to the observed decline in relapse rates. When trying to establish a causal relationship, the (a priori) plausibility should eventually also come into consideration. This understanding will allow us to make rational comparisons of therapeutic effects as increasing numbers of therapies for RMS emerge. In particular, this highlights the importance of considering certain variables, like population characteristics, pre-trial ARR and the particularities of its measurement already in the design of studies as well as when reporting and interpreting study results.

**Acknowledgements**

The authors would like to thank two anonymous referees for the insightful comments. RN is grateful for support from the NIHR Biomedical Research Centre.

**Table 1: Baseline characteristics of all randomized patients and the placebo groups in the 56 randomized, controlled trials included in the systematic review.**

|  | Placebo patients | | All randomized patients | |
|---|---|---|---|---|
|  | N | Median (range) | N | Median (range) |
| **Number of patients** | 55 | 54 (7 - 556) | 55 | 148 (13-1644) |
| **Percentage female** | 47 | 68.1 (41.2 - 82.9) | 48 | 69.6 (52.6-82.8) |
| **Mean age (years)** | 46 | 36.7 (26.5 – 43.0) | 46 | 36.0 (27.7 – 43.6) |
| **Mean MS duration (years)** | 40 | 7.2 (2.1 – 11.0) | 40 | 7.1 (2.6 – 11.7) |

**Figure 1: Pre-trial annualized relapse rate (ARR) and on-trial placebo ARR observed in the 56 trials identified by our literature search against the calendar year in which the papers were published. The size of the circles indicates the size of the trial and is inversely proportional to the standard error of the ARR. The solid lines show the model values and the dashed lines indicate 95% confidence intervals.**

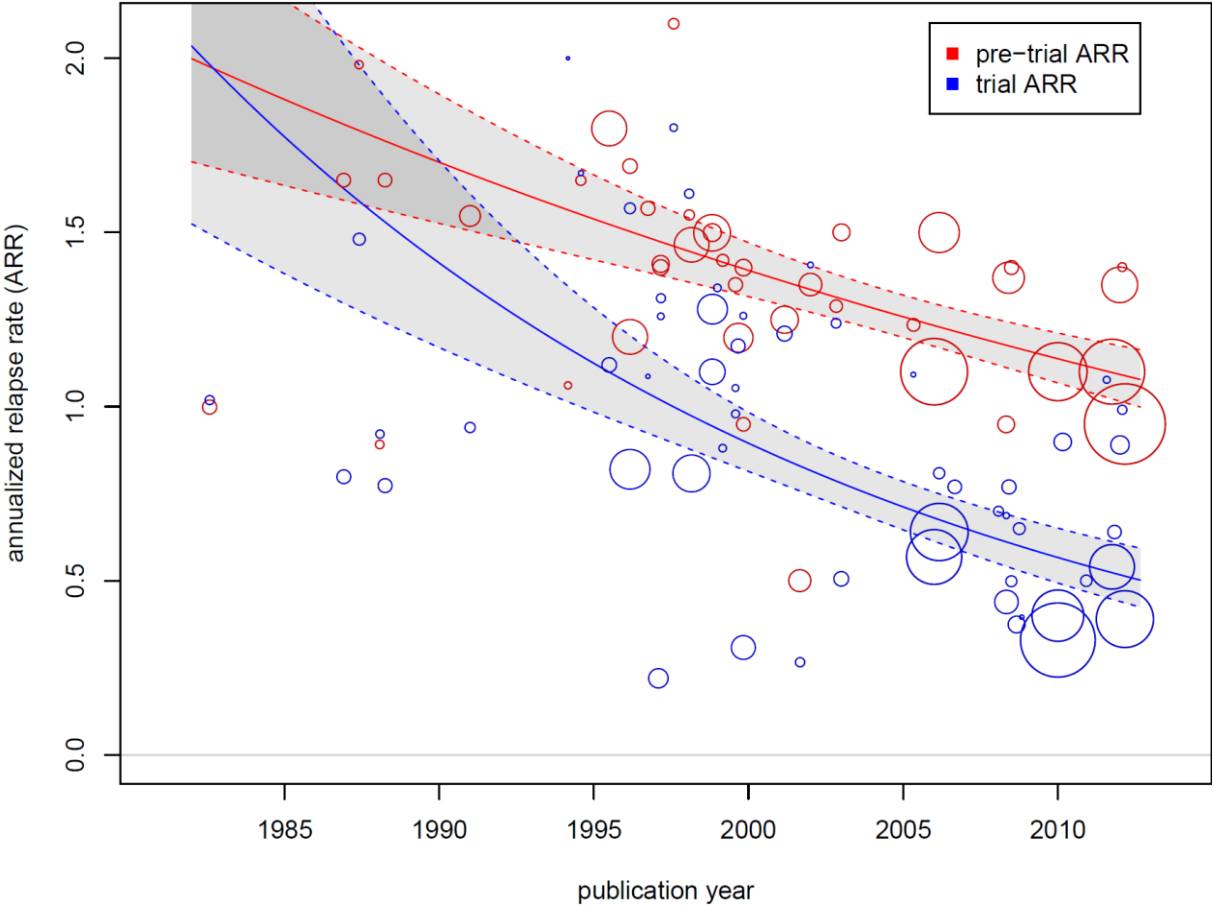

**Figure 2: Rate ratio of ARRs calculated for the year before inclusion and the year before that for trials reporting ARR for 12 and 24 months prior to the study.**

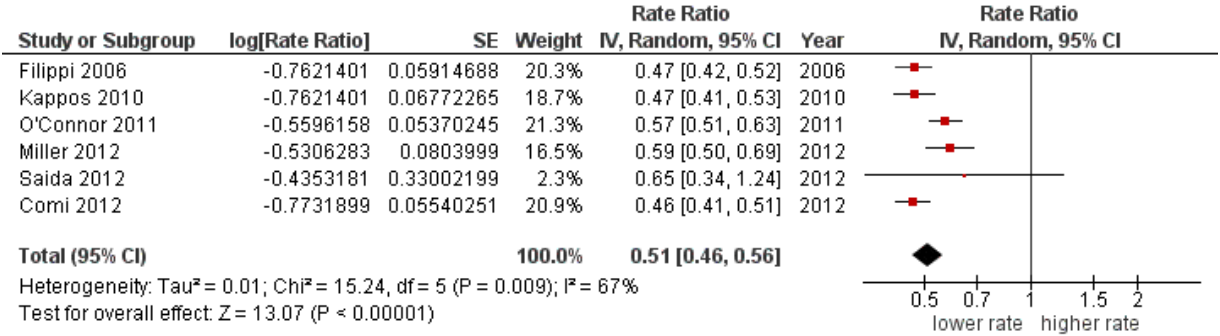

IV: Inverse variance weights; CI: Confidence interval; SE: Standard error

**Figure 3: Mean duration of illness of placebo patients in years, weighted by the inverse standard error squared. Whiskers indicate the doubled standard error, if available. The linear fit takes into account all values with standard errors – the remaining items are colored grey.**

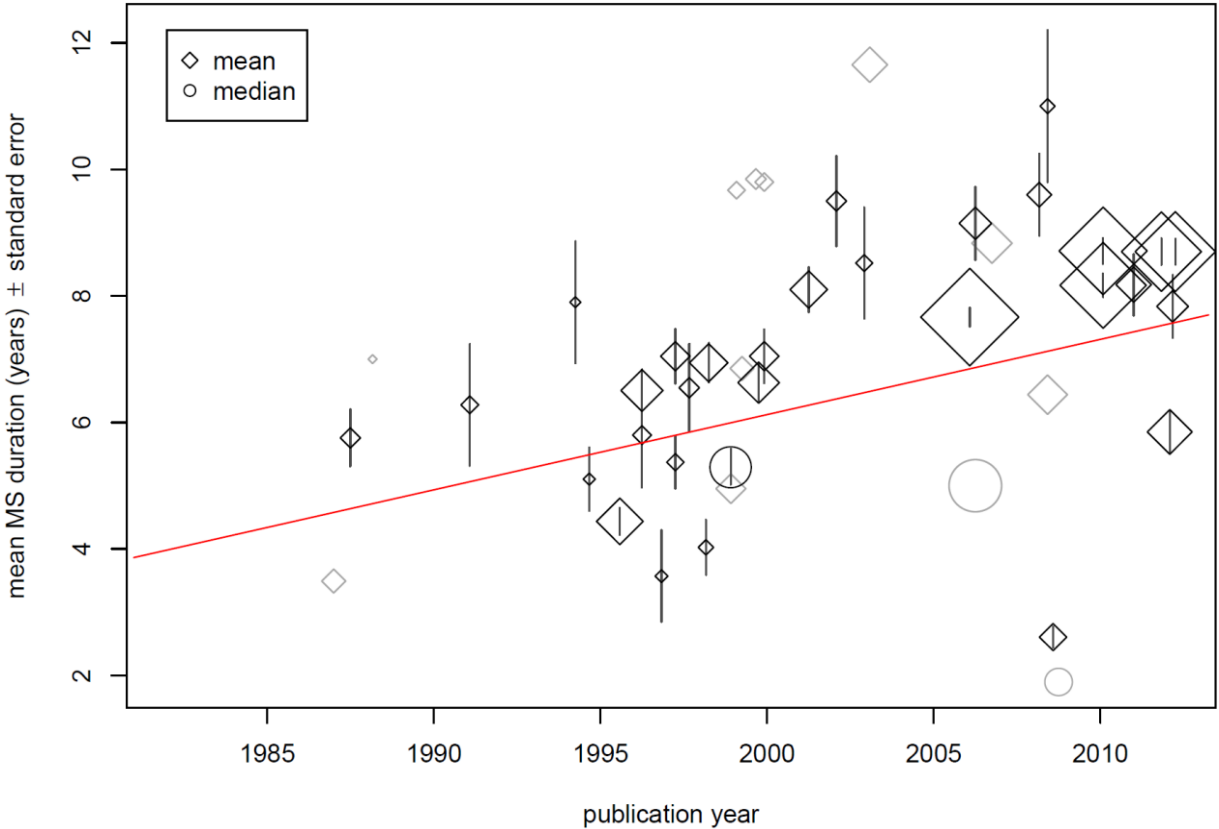

**Figure 4: Factors driving the temporal trend in placebo ARR and their importance.** Left column: The temporal trend line for the (logarithmic) trial ARR shown in Figure 1 explains about 46% of the observed variation in trial ARR over the years. Right column: The Meta-regression explains about 69% of the variation in trial ARR; the temporal trend is insignificant with major contributors being pre-trial ARR, the number of years used to calculate pre-trial ARR, study duration, and MS duration.

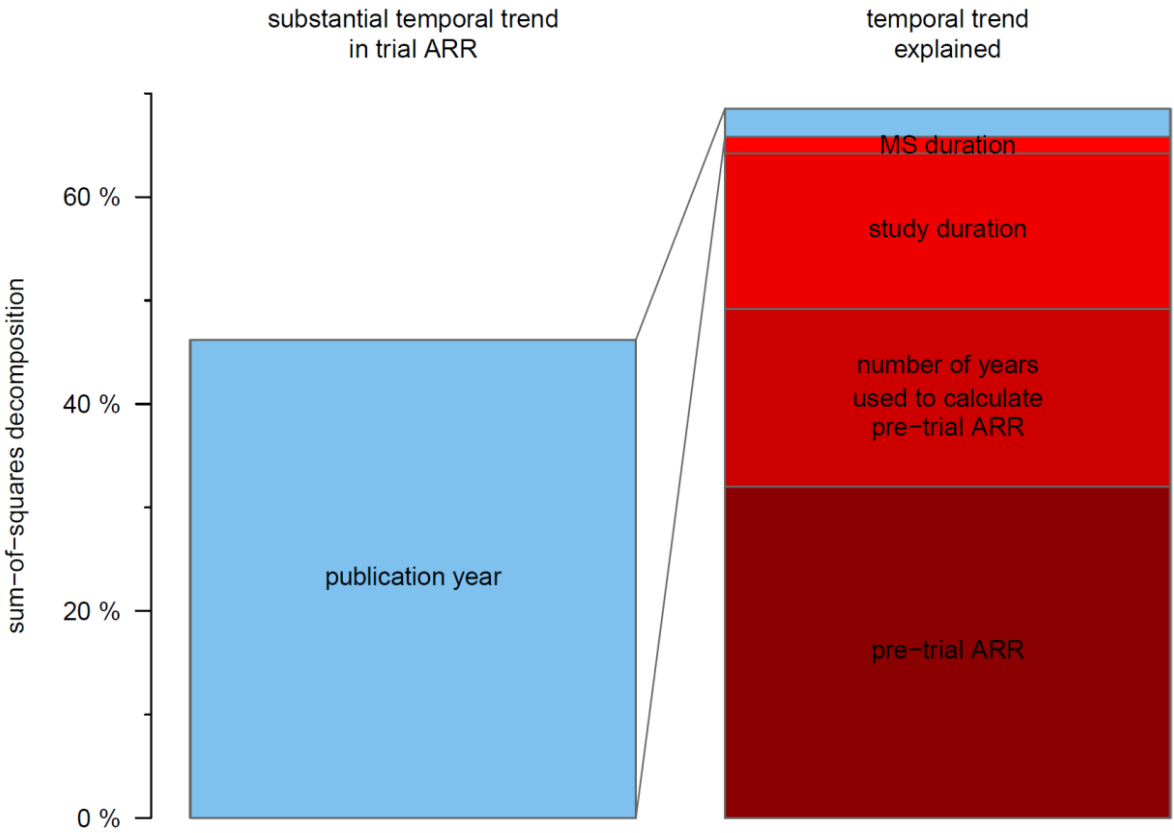

# Explaining temporal trends in annualized relapse rates in placebo groups of randomized controlled trials in relapsing multiple sclerosis: systematic review and meta-regression

Simon M Steinvorth, Christian Röver, Simon Schneider, Richard Nicholas, Sebastian Straube, Tim Friede

**WEB APPENDIX**

**Literature search**

Our systematic literature search resulted in the identification of the 56 randomized, placebo-controlled trials in relapsing multiple sclerosis reported in the following publications (in order of publication):

[A1]   Gonsette RE, Demonty L, Delmotte P et al. Modulation of immunity in multiple sclerosis: a double-blind levamisole-placebo controlled study in 85 patients. *J. Neurol.* 1982; 228: 65-72.

[A2]   Mertin J, Rudge P, Kremer M et al. Double-blind controlled trial of immunosuppression in the treatment of multiple sclerosis: final report. *The Lancet* 1982; 320: 351-354.

[A3]   Camenga DL, Johnson KP, Alter M et al. Systemic recombinant alpha-2 interferon therapy in relapsing multiple sclerosis. *Archives of Neurology* 1986; 43: 1238.

[A4]   Jacobs L, Salazar AM, Herndon R et al. Intrathecally Administered Natural Human Fibroblast Interferon Reduces Exacerbations of Multiple Sclerosis. *Archives*

**Table A1: Sizes and durations of the individual studies considered here.**

| | study | duration (years) | total patients | placebo patients |
|---|---|---|---|---|
| [A1] | Gonsette, R. E. et al. (1982) | 2.00 | | |
| [A2] | Mertin, J. et al. (1982) | 1.25 | 43 | 22 |
| [A3] | Camenga, D. L. et al. (1986) | 1.00 | 98 | 50 |
| [A4,A5] | Jacobs, L. et al. (1987) | 2.00 | 69 | 35 |
| [A6] | Hirsch, R. L. et al. (1988) | 1.00 | 98 | 50 |
| [A7] | Milanese, C. et al. (1988) | 3.00 | 13 | 7 |
| [A8] | Goodkin, D. E. et al. (1991) | 2.00 | 54 | 25 |
| [A9] | Bastianello, S. et al. (1994) | 1.00 | 25 | 12 |
| [A10] | Durelli, L. et al. (1994) | 0.50 | 20 | 8 |
| [A11,A12] | IFNB (1995) | 5.00 | 372 | 123 |
| [A13] | Andersen, O. et al. (1996) | 0.46 | 28 | 14 |
| [A14] | Jacobs, L. D. et al. (1996) | 2.99 | 301 | 143 |
| [A15] | Lycke, J. et al. (1996) | 2.00 | 60 | 30 |
| [A16] | Fazekas, F. et al. (1997) | 2.00 | 148 | 73 |
| [A17] | Millefiorini, E. et al. (1997) | 2.00 | 51 | 24 |
| [A18] | Miller, A. E. et al. (1997) | 0.50 | 103 | 54 |
| [A19,A20] | Van Oosten, B. W. et al. (1997) | 1.50 | 71 | 36 |
| [A21] | Achiron, A. et al. (1998) | 2.00 | 40 | 20 |
| [A22,A23] | Johnson, K. P. et al. (1998) | 2.92 | 251 | 126 |
| [A24] | Noseworthy, J. H. et al. (1998) | 3.00 | 151 | 79 |
| [A25] | PRISMS (1998) | 2.00 | 560 | 187 |
| [A26] | Deisenhammer, F. et al. (1999) | 2.00 | 148 | 73 |
| [A27] | Lenercept (1999) | 0.46 | 168 | 44 |
| [A28] | Myhr, K. M. et al. (1999) | 0.50 | 97 | 33 |
| [A29] | OWIMS (1999) | 0.92 | 293 | 100 |
| [A30] | Patti, F. et al. (1999) | 2.00 | 58 | 29 |
| [A31] | Romine, J. S. et al. (1999) | 1.50 | 52 | 25 |
| [A32] | Tubridy, N. et al. (1999) | 0.46 | 72 | 35 |
| [A33] | Brod, S. A. et al. (2001) | 0.75 | 29 | 10 |
| [A34] | Comi, G. et al. (2001) | 0.75 | 239 | 120 |
| [A35] | Bech, E. et al. (2002) | 0.46 | 70 | 34 |
| [A36] | Lewanska, M. et al. (2002) | 1.00 | 49 | 17 |
| [A37] | Miller, D. H. et al. (2003) | 0.50 | 213 | 71 |
| [A38] | Wroe, S. J. (2005) | 0.25 | 98 | 33 |
| [A39] | Filippi, M. et al. (2006) | 1.07 | 1644 | 548 |
| [A40] | Kappos, L. et al. (2006) | 0.50 | 277 | 92 |
| [A41] | O'Connor, P. W. et al. (2006) | 0.69 | 179 | 61 |
| [A42] | Polman, C. H. et al. (2006) | 2.30 | 942 | 315 |
| [A43] | Broadley, S. A. et al. (2008) | 0.23 | 50 | 11 |
| [A44] | Comi, G. et al. (2008) | 0.69 | 306 | 102 |
| [A45] | Fazekas, F. et al. (2008) | 0.92 | 127 | 41 |
| [A46] | Garren, H. et al. (2008) | 0.92 | 267 | 87 |
| [A47] | Hauser, S. L. et al. (2008) | 0.92 | 104 | 35 |
| [A48] | Kappos, L. et al. (2008) | 0.46 | 256 | 65 |
| [A49] | Mostert, J. P. et al. (2008) | 0.46 | 38 | 19 |
| [A50] | Segal, B. M. et al. (2008) | 0.71 | 249 | 49 |
| [A51] | Barkhof, F. et al. (2010) | 1.00 | 292 | 100 |
| [A52] | Giovannoni, G. et al. (2010) | 1.84 | 1326 | 437 |
| [A53] | Kappos, L. et al. (2010) | 2.00 | 1272 | 418 |
| [A54] | Vollmer, T. L. et al. (2010) | 0.46 | 215 | 69 |
| [A55,A56] | De Stefano, L. et al. (2011) | 0.31 | 180 | 60 |
| [A57] | Kappos, L. et al. (2011) | 0.46 | 218 | 54 |
| [A58] | O'Connor, P. et al. (2011) | 2.07 | 1088 | 363 |
| [A59] | Comi, G. et al. (2012) | 2.00 | 1106 | 556 |

| [A60] | Miller, D. H. et al. (2012) | 0.46 | 343 | 99 |
| [A61] | Saida, T. et al. (2012) | 0.50 | 171 | 57 |

**Table A2: Results of regression analyses to check for time trends in individual variables. The percentage of trials for which the corresponding data are available is shown, as well as the value , confidence interval and p-value for the regression coefficient indicating the yearly change.**

| Variable | available data | p-value | coefficient | 95% confidence interval |
|---:|:---:|:---:|:---:|:---:|
| trial ARR* | 100% | **<0.001** | -0.032 | [-0.042, -0.022] |
| log (trial ARR)* | 100% | **<0.001** | -0.046 | [-0.059, -0.032] |
| number of years used to calculate pre-trial ARR | 75% | **<0.001** | -0.049 | [-0.071, -0.027] |
| minimum ARR at inclusion | 86% | 0.581 | 0.003 | [-0.008, 0.013] |
| minimum relapse-free days at inclusion | 55% | 0.247 | -0.623 | [-1.701, 0.456] |
| minimum steroid-free days at inclusion | 68% | 0.059 | -0.991 | [-2.019, 0.038] |
| minimum EDSS at inclusion | 95% | 0.051 | -0.028 | [-0.056, 0.000] |
| maximum EDSS at inclusion | 95% | 0.986 | -0.0003 | [-0.036, 0.035] |
| eligibility criteria: number | 100% | **<0.001** | 0.771 | [0.392, 1.151] |
| eligibility criteria: words | 100% | **<0.001** | 6.651 | [3.372, 9.930] |
| eligibility criteria: characters | 100% | **<0.001** | 39.638 | [20.109, 59.168] |
| Oxford quality scale | 100% | **0.021** | 0.028 | [0.004, 0.051] |
| baseline age (years)* | 82% | **<0.001** | 0.198 | [0.098, 0.299] |
| pre-trial ARR* | 73% | **<0.001** | -0.027 | [-0.039, -0.015] |
| log(pre-trial ARR)* | 73% | 0.097 | -0.020 | [-0.044, 0.004] |
| baseline EDSS* | 79% | 0.289 | 0.011 | [-0.010, 0.031] |
| baseline MS duration (years)* | 82% | **0.048** | 0.122 | [0.001, 0.243] |
| baseline number of patients | 98% | **0.004** | 6.664 | [2.270, 11.059] |
| percentage of females* | 84% | 0.337 | -0.122 | [-0.004, 0.001] |
| study duration (years) | 100% | **0.006** | -0.043 | [-0.074, -0.013] |
| patient-years | 98% | 0.060 | 8.853 | [-0.386, 18.092] |
| number of treatment arms | 100% | **<0.001** | 0.048 | [0.027, 0.069] |
| mean number of patients per treatment arm | 98% | **0.003** | 7.028 | [2.426, 11.630] |

* weighted linear regression (based on standard errors)

**statistically significant (p < 0.05)**

**Table A3: Results of the sub-group analysis; shown is the estimated mean and standard error for each variable within the given cluster of studies as defined by period of publication. The p-values indicate whether there is a significant difference among clusters.**

| Variable | p-value | Cluster I (...-1994) | Cluster II (1995-2000) | Cluster III (2001-2009) | Cluster IV (2010-...) |
|---|---|---|---|---|---|
| trial ARR* | **<0.001** | 1.15 (0.34) | 1.29 (0.14) | 0.76 (0.14) | 0.42 (0.03) |
| log(trial ARR)* | **<0.001** | 0.29 (0.33) | 0.29 (0.13) | -0.17 (0.22) | -0.70 (0.09) |
| number of years used to calculate pre-trial ARR | **0.008** | 2.21 (0.20) | 1.97 (0.13) | 1.42 (0.16) | 1.50 (0.24) |
| minimum ARR at inclusion | 0.824 | 1.00 (0.10) | 1.00 (0.06) | 0.96 (0.07) | 1.07 (0.09) |
| minimum relapse-free days at inclusion | 0.341 | 60.9 (12.5) | 38.2 (5.6) | 39.2 (4.9) | 34.3 (7.2) |
| minimum steroid-free days at inclusion | **0.020** | 68.5 (9.5) | 35.5 (5.1) | 38.4 (5.3) | 31.6 (7.2) |
| minimum EDSS at inclusion | 0.107 | 0.67 (0.29) | 0.63 (0.16) | 0.17 (0.17) | 0.10 (0.23) |
| maximum EDSS at inclusion | 0.269 | 6.00 (0.36) | 5.24 (0.20) | 5.47 (0.21) | 5.70 (0.28) |
| eligibility criteria: number | **0.004** | 8.9 (3.8) | 13.8 (2.6) | 18.9 (2.7) | 27.3 (3.6) |
| eligibility criteria: words | **0.002** | 82 (32) | 126 (22) | 158 (23) | 253 (30) |
| eligibility criteria: characters | **0.002** | 459 (190) | 711 (131) | 894 (135) | 1486 (181) |
| Oxford quality scale | 0.291 | 4.00 (0.23) | 4.42 (0.16) | 4.44 (0.16) | 4.60 (0.22) |
| baseline age (years)* | **0.004** | 32.0 (2.3) | 35.4 (0.6) | 36.5 (0.5) | 38.0 (0.5) |
| pre-trial ARR* | **0.001** | 1.57 (0.26) | 1.47 (0.08) | 1.20 (0.07) | 1.06 (0.05) |
| log(pre-trial ARR)* | **0.014** | 0.50 (0.24) | 0.48 (0.08) | 0.02 (0.10) | 0.20 (0.19) |
| baseline EDSS* | **0.004** | 3.67 (0.48) | 2.47 (0.10) | 2.35 (0.09) | 2.72 (0.08) |
| baseline MS duration (years)* | **0.007** | 5.84 (1.73) | 5.15 (0.61) | 5.45 (0.64) | 8.15 (0.61) |
| baseline number of patients | **0.002** | 52 (114) | 159 (74) | 285 (76) | 621 (102) |
| percentage of females* | 0.340 | 65.6 (4.8) | 72.1 (1.5) | 69.7 (1.4) | 69.2 (1.1) |
| study duration (years) | **0.004** | 1.53 (0.29) | 1.80 (0.20) | 0.75 (0.20) | 1.11 (0.27) |
| patient-years | **0.018** | 36 (86) | 147 (56) | 104 (57) | 375 (77) |
| number of treatment arms | **0.001** | 2.00 (0.21) | 2.32 (0.14) | 2.83 (0.15) | 3.00 (0.20) |
| mean number of patients per treatment arm | **0.005** | 26 (44) | 66 (28) | 106 (29) | 224 (39) |

\* weighted linear regression (based on standard errors)

**statistically significant ($p < 0.05$)**